      [1999/12/01 v1.4c Il Nuovo Cimento]
\documentclass{cimento}
\usepackage{graphicx}  
\title{Kaon physics with KLOE}
\author{B.~Sciascia\from{ins:x}\thanks{For the KLOE Collaboration:
F.~Ambrosino, A.~Antonelli, M.~Antonelli, F.~Archilli, P.~Beltrame, 
G.~Bencivenni, C.~Bini, C.~Bloise, S.~Bocchetta, F.~Bossi, P.~Branchini, 
G.~Capon, D.~Capriotti, T.~Capussela, F.~Ceradini, P.~Ciambrone, E.~De Lucia, 
A.~De Santis, P.~De Simone, G.~De Zorzi, A.~Denig, A.~Di Domenico, C.~Di Donato, 
B.~Di Micco, M.~Dreucci, G.~Felici, S.~Fiore, P.~Franzini, C.~Gatti, P.~Gauzzi, 
S.~Giovannella, E.~Graziani, M.~Jacewicz, V.~Kulikov, 
J.~Lee-Franzini, 
M.~Martini, P.~Massarotti, S.~Meola, S.~Miscetti, M.~Moulson, S.~M\"uller, F.~Murtas, 
M.~Napolitano, F.~Nguyen, M.~Palutan, A.~Passeri, V.~Patera, P.~Santangelo, B.~Sciascia, 
A.~Sibidanov, T.~Spadaro, 
C.~Taccini, L.~Tortora, P.~Valente, G.~Venanzoni, and R.~Versaci.}}
\instlist{\inst{ins:x} Laboratori Nazionali di Frascati - INFN }
\PACSes{\PACSit{13.20}{Eb}
\PACSit{12.15}{Hh}}
\begin{document}

\maketitle

\begin{abstract}
Kaon physics can test new-physics effects in leptonic or semileptonic decays.
A unitarity test of the first row of the CKM mixing matrix is obtained
from the precision measurements of Kl3 widths for $K^\pm$, $K_L$ , and (unique to KLOE) $K_S$.
The KLOE measurement of $R_K = \Gamma(Ke2)/\Gamma(K\mu2)$ with an accuracy at the \% level, 
aims at finding evidence of deviations from the SM prediction induced by lepton-flavor violation 
new-physics effects.
\end{abstract}

\newcommand{\Vud}{\ensuremath{|V_{ud}|}}
\newcommand{\Vus}{\ensuremath{|V_{us}|}}
\newcommand{\Vusf}{\ensuremath{|V_{us}|f_+(0)}}
\newcommand{\abs}[1]{\ensuremath{\left|#1\right|}}
\newcommand{\kpm} {\ensuremath{K^{\pm}}}
\newcommand{\kp} {\ensuremath{K^{+}}}
\newcommand{\pipm} {\ensuremath{\pi^{\pm}}}
\newcommand{\vub} {\ensuremath{|V_\mathrm{ub}|}}
\newcommand{\fzero}  {\ensuremath{f_+(0)}}
\newcommand{\fzerokpi}  {\ensuremath{f_+^{K^0\pi^-}(0)}}
\newcommand{\kzero} {\ensuremath{K^{0}}}
\newcommand{\kltre}  {\ensuremath{K_{l3}}}
\newcommand{\kletre}  {\ensuremath{K_{L}e3}}
\newcommand{\ksetre}  {\ensuremath{K_{S}e3}}
\newcommand{\klmutre}  {\ensuremath{K_{L}\mu3}}
\newcommand{\kpmetre}   {\ensuremath{K^{\pm} {e3}}}
\newcommand{\kpmmutre}  {\ensuremath{K^{\pm} {\mu3}}}
\newcommand{\pippimpio}  {\ensuremath{\pi^+\pi^-\pi^0}}
\newcommand{\pippim}  {\ensuremath{\pi^+\pi^-}}
\newcommand{\duepio}  {\ensuremath{2\pi^0}}
\newcommand{\trepio}  {\ensuremath{3\pi^0}}
\newcommand{\mudue}  {\ensuremath{\mu^{\pm}\nu}}
\newcommand{\taus}  {\ensuremath{ \tau_S}}
\newcommand{\taul}  {\ensuremath{\tau_L}}
\newcommand{\taupm}  {\ensuremath{\tau_{\pm}}}
\newcommand{\kppipigall}{\ensuremath{K^{+} \rightarrow \pi^{+}\pi^0\,(\gamma)}}
\let\kpppg=\kppipigall
\def\ff{$\phi-$factory}
\def\up#1{$^{#1}$}  \def\dn#1{$_{#1}$}
\def\pt#1,#2,{\ifm{#1\x10^{#2}}}
\newcommand{\ka}[1]{\ensuremath{K^{{#1}}}}
\newcommand{\lam}[2]{\ensuremath{\lambda^{{#1}}_{{#2}}}}
\newcommand{\eV}{{e\kern-.07em V}}
\newcommand{\MeV}{{\rm \,M\eV}}
\hyphenation{Flavia-Net}
\newcommand{\mtaud} {m^2_{\tau}}
\newcommand{\mkd} {m^2_K}
\newcommand{\mkq} {m^4_K}
\newcommand{\mpid} {m^2_\pi}
\def\ni {\noindent}
\def\rd{\mbox{d}}
\def\IK{I_{K_{e3}^0}}
\def\eqalign#1{\null\,\vcenter{\openup\jot\m@th
  \ialign{\strut\hfil$\displaystyle{##}$&$\displaystyle{{}##}$\hfil
      \crcr#1\crcr}}\,}
\newcommand{\szero}{{\it step-0}}
\newcommand{\suno}{{\it step-1}}
\newcommand{\text}{\rm}
\newcommand{\BR}{\ensuremath{\mathcal{B}}}
\newcommand{\nc}{\newcommand}
\nc{\am}[1]{\ensuremath{#1}}

\nc{\ke}{\am{K \to e \nu}}
\nc{\km}{\am{K \to \mu \nu}}
\nc{\kpd}{\am{K_{\pi 2}}}
\nc{\ked}{\am{K_{e2}}}
\nc{\kedg}{\am{K_{e2\gamma}}}
\nc{\ket}{\am{K_{e3}}}
\nc{\kmt}{\am{K_{\mu3}}}

\nc{\kmd}{\am{K_{\mu2}}}
\nc{\erreg}{\am{R_{\gamma}}}
\nc{\kee}{\am{K \to \pi  e \nu}}
\nc{\kmm}{\am{K \to \pi \mu \nu}}
\nc{\nn}{\am{N\!N}}
\nc{\NN}{\am{N\!N}}

\nc{\be}{\begin{equation}}
\nc{\ee}{\end{equation}}
\nc{\ChPT}{$\chi$PT}
\nc{\GeV}{\mbox{GeV}}
\nc{\ps}{\mbox{ps}}
\nc{\mrad}{\mbox{mrad}}
\nc{\ie}{i.e.}
\nc{\mbo}{\mathversion{bold}}
\def\cl{\centerline} \def\f{$\phi$}  \def\ab{$\sim$}  \def\dif{\hbox{d}}
\def\red{\color{red}}  \def\blue{\color{blue}}  \def\gam{\gamma}
\def\vp{{\vphantom{$I^{I^i}$}}}

\section{Introduction}
Purely leptonic and semileptonic  decays of K mesons ($K  \to \ell \nu, K \to   \pi \ell \nu$, $\ell = e, \mu$)  
are mediated in the Standard Model (SM) by tree-level W-boson exchange. 
Gauge coupling universality and three-generation  quark mixing imply that 
semileptonic processes such as $d^i \to  u^j \ell \nu$ are governed by 
the effective Fermi constant $G_{ij} = G_\mu \, V_{ij}$, 
where $G_\mu$ is the  muon decay constant and
(ii) $V_{ij}$ are the elements of the unitary  Cabibbo--Kobayashi 
Maskawa (CKM) matrix. This fact has simple but deep consequences, 
that go under the name of universality relations.
In the SM  the effective semileptonic 
constant $G_{ij}$ does not depend on the lepton flavor. 
If one extracts $V_{ij}$ from different semileptonic transitions assuming 
quark-lepton gauge universality (i.e. normalizing the decay rates with $G_\mu$), the CKM unitarity 
condition $\sum_j  |V_{ij}|^2 = 1$ should be verified. 

Beyond the SM,  these universality relations can be violated 
by new  contributions 
to the low-energy  $V$-$A$  four fermion operators,  
as well as new non $V$-$A$ structures.  
Therefore,  precision tests of the universality relations 
probe physics beyond the SM and are sensitive to several 
SM extensions~\cite{Marciano:1987ja,Hagiwara:1995fx,Kurylov:2001zx,Cirigliano:2009wk}.

This paper is organized as follows. The present and future status of DA\char8NE accelerator and
KLOE experiment is briefly reviewed in Sect. \ref{DeK}. The world average measurement of $V_{us}$ is presented in 
Sect. \ref{Vus} together with the new preliminary KLOE measurements of $K_L$ and $K_S$ lifetimes.
The KLOE result for $R_K$ is described in Sect. \ref{Kedue}.

\section{DA\char8NE and KLOE: present and future}
\label{DeK}
DA\char8NE, the Frascati $\phi$-factory, is an $e^+e^-$ collider working at 
$\sqrt{s}\sim m_\phi\sim1.02$\GeV. $\phi$ mesons are produced, essentially at rest, 
with a visible cross section of $\sim$3.1 $\mu$b.
During year 2008 the Accelerator Division 
has tested a new interaction scheme 
with the goal of reaching a peak luminosity of 5$\times$10$^{32}$ cm$^{-2}$s$^{-1}$,
a factor of three larger than what previously obtained.

KLOE is a multipurpose detector, mainly consisting of a large cylindrical drift chamber (DC) 
with an internal
radius of 25 cm and an external one of 2 m, surrounded by
a lead-scintillating fibers electromagnetic calorimeter (EMC).
Both are immersed in the 0.52 T field of a superconducting solenoid.
From 2000 to 2006, KLOE has acquired 2.5 fb$^{-1}$ of data
at the $\phi$(1020) peak, plus additional 250 pb$^{-1}$ at
energies slightly higher or lower than that. 
A collection of the main physics results of KLOE and details of the detector can be found in 
Ref. \cite{Bossi:2008aa} and references therein. 

For the forthcoming run \cite{KLOE2LoI}, 
upgrades have also been proposed for the detector.
In a first phase, two different devices 
will be installed along the beam line to detect the scattered electrons/positrons
from $\gamma\gamma$ interactions.
In a second phase, a light--material internal tracker 
will be installed
in the region between the beam pipe and the drift chamber to improve charged vertex
reconstruction and to increase the acceptance for low p$_{T}$ 
tracks. 
Crystal calorimeters 
will cover the low $\theta$ region, aiming at
increasing acceptance for very forward electrons/photons down to 
8$^\circ$. A new tile calorimeter 
will be used to instrument 
the DA\char8NE focusing system for the detection of photons coming from $K_L$  
decays in the drift chamber. Implementation of the second phase is planned for late 2011.
The integrated luminosity for the two phases
should be 5 fb$^{-1}$ and 20 fb$^{-1}$, respectively. 

\section{Measurement of $V_{us}$}
\label{Vus}

Large amount of data  has been collected on the semileptonic 
modes $K \to \pi \ell \nu$ by several experiments, BNL-E865, KLOE, KTeV, ISTRA+, and NA48 
in the last few years.
These data have stimulated a sub\-stan\-ti\-al pro\-gress on the theoretical 
inputs, so that most of the theory-do\-mi\-na\-ted errors associated to radiative 
corrections 
and hadronic form factors 
have been reduced below $1\%$.
Presently,  the unitarity test
\begin{equation}
|V_{ud}|^2 + |V_{us}|^2 +|V_{ub}|^2 = 1 + \Delta_{\rm CKM}
\label{eq:unitarity}
\end{equation}
implies that $\Delta_{\rm CKM}$ is consistent with zero at the level of $6 \times 10^{-4}$.  
$V_{us}$ from $K \to \pi \ell \nu$ decays contributes about half of this uncertainty, 
mostly coming from the hadronic matrix element. 
Both experimental and the\-o\-re\-ti\-cal pro\-gress in $K_{\ell 3}$  decays 
will be needed in order to improve the accuracy on $\Delta_{\rm CKM}$ in the future.  

It has been shown~\cite{Cirigliano:2009wk} that presently semileptonic processes and the related 
universality tests provide constraints on NP that cannot be obtained from other electroweak
precision tests and/or direct measurements at the colliders.

In the last years, many efforts have been dedicated to the correct averaging of the rich
harvest of recent results in kaon physics. The FLAVIAnet kaon working group 
has published a comprehensive review~\cite{Antonelli:2008jg} in 2008
where a detailed description of the averaging procedure can be found.
However, the significant progress on both the experimental and theoretical sides,
has motivated the same group to publish an updated analysis~\cite{Antonelli:2010aa}.
Even if these proceedings 
will focus on the contribution from KLOE,  
all the $V_{us}$-related results presented refer to the FlaviaNet working group outcomes.

After four years of data analysis, KLOE has produced the most comprehensive set of results
from a single experiment, measuring the main 
BRs of $K_L$, 
$K^{\pm}$, 
and $K_S$ 
(unique to KLOE), including semileptonic and two-body decays;
lifetime measurements for $K_L$ 
and $K^\pm$, 
form factor slopes from the analysis of $K_{L}e3$ 
and $K_{L}\mu3$. 
The value of \Vusf\ 
has been obtained from KLOE results \cite{Ambrosino:2008ct} 
using the $K_S$ lifetime
from PDG \cite{Amsler:2008zzb} as the only non--KLOE input.
The values of \Vusf\ obtained from the world average of K semileptonic measurements~\cite{Antonelli:2010aa} 
are shown in Tab. \ref{tab:vusf}. 
\begin{table}[!ht]
\begin{center}
\caption{\label{tab:vusf}Values of \Vusf\  extracted from \kltre\ decay rates.}
\begin{tabular}{c c c c c} 
\hline
 $K_Le3$  &  $K_L\mu3$ & $K_Se3$    & $K^{\pm}e3$ & $K^{\pm}\mu3$ \\ \hline
 0.2163(6)&  0.2166(6)& 0.2155(13) & 0.2160(11) & 0.2158(14)  \\ \hline
\end{tabular}
\end{center}
\end{table}

The five decay modes agree well within the errors and average to \Vusf\ = 0.2163(5), with 
$\chi^2/ndf =0.77/4$ (Prob=94\%).
Significant lepton-universality tests are provided by the comparison of 
the results from different leptonic channels. 
Defining the ratio $r_{\mu e}=\Vusf_{\mu3}^2/\Vusf_{e3}^2$
we have $r_{\mu e} = g_{\mu}^2/ g_{e}^2$,
with $g_{\ell}$ the coupling strength at the $W \to \ell \nu$ vertex.
Lepton universality can be then tested comparing the measured value 
of $r_{\mu e}$ with the SM prediction $r_{\mu e}^{SM}=1$. 
Averaging charged- and neutral-kaon modes, 
we obtain $r_{\mu e} =1.002(5)$, to be compared with the results  
from leptonic pion decays,
$(r_{\mu e})_{\pi} =1.0042(33)$ \cite{RamseyMusolf:2007yb}, and from 
leptonic $\tau$ decays 
$(r_{\mu e})_{\tau} = 1.000(4)$ \cite{Davier:2005xq}. 

Using the determination of \Vusf\  from \kltre\ decays and 
the value $\fzero = 0.959(5)$ (see Ref. \cite{Antonelli:2010aa} for a detailed discussion on this choice),
we get $\Vus = 0.2254(13)$.
 
Furthermore, a measurement of $\Vus/\Vud$ can be obtained from the  
comparison of the radiation-inclusive decay rates of $\kpm \rightarrow \mudue (\gamma)$ 
and $\pipm \rightarrow \mudue (\gamma)$, combined with lattice calculation of $f_K/f_{\pi}$ \cite{Marciano:2004uf}.
Using the BR$(\kpm \rightarrow \mudue)$ average value (dominated by KLOE result \cite{Ambrosino:2005fw})
and the lattice result $f_K/f_\pi = 1.193(6)$ 
(again see Ref. \cite{Antonelli:2010aa} for a detailed discussion on this choice),
we get $\Vus/\Vud = 0.2312(13)$. 
This value can be used in a fit together with the measurements of
\Vus\ from \kltre\ decays and $\Vud = 0.97425(22)$ \cite{TownerHardy} 
from superallowed nuclear $\beta$ decays.
The result of this fit is $\Vud = 0.97425(22)$ and $\Vus = 0.2253(9)$, with 
$\chi^2/ndf =0.014/1$ (Prob$ = 91\%$), 
from which we get $1-(\Vud^2+\Vus^2+\vub^2)=-0.0001(6)$ which is in striking agreement with the unitarity hypothesis.
Using these results, we evaluate 
$G_{\rm CKM}=G_\mu \sqrt{\Vud^2 + \Vus^2+\vub^2}=1.16633(35)\times 10^{-5}$ GeV$^{-2}$,
with $G_\mu=1.166371(6)\times 10^{-5}$ GeV$^{-2}$.
At present, the sensitivity of the quark--lepton universality test 
through the $G_{\rm CKM}$ measurement is competitive and even 
better than  
the measurements from $\tau$ decays and the electroweak precision 
tests \cite{Marciano:2007zz}.  
Thus unitarity can also be interpreted as a test of 
the universality of lepton and quark weak couplings to the $W$ boson,  
allowing bounds to be set on extensions of the SM leading to some kind of 
universality breaking.
For instance, the existence of additional $Z^\prime$ gauge bosons, 
giving different loop-contributions to muon and 
semileptonic decays, can break gauge universality
\cite{Marciano:1987ja}. 
The measurement of $G_{\rm CKM}$ can set constraints on 
the $Z^\prime$ mass which
are competitive with direct search at the colliders. 
When considering supersymmetric extensions, differences between muon 
and semileptonic decays can arise in the loop 
contributions from SUSY particles~\cite{Hagiwara:1995fx,Kurylov:2001zx}.  
The slepton-squark mass difference could be investigated improving 
present accuracy on the unitarity relation by a factor of $\sim$2-3. 

\subsection{$K_L$ lifetime}
The error on the $K_L$ lifetime ($\tau_L$) determination is the limiting factor on \Vusf\ when calculated from $K_L$.
Using all the available data, KLOE can improve statistical and systematic error over its previous 
measurements~\cite{Ambrosino:2005vx,Ambrosino:2005ec}.
KLOE decided to perform a new $\tau_L$ measurement based on 46 million $K_L\to3\pi^0$ events; using the 
same method used in Ref. \cite{Ambrosino:2005vx}.
$K_L$ mesons are tagged by detecting $K_S\to\pi^+\pi^-$ decays and the time dependence of the $K_L\to3\pi^0$
decays is used to measure the $K_L$ lifetime. The $3\pi^0$ mode is chosen because is the most
abundant, has high detection efficiency and the tagging efficiency is almost independent of the $K_L$
path length. 
The preliminary result is: $\tau_L = 50.56 \pm 0.14_{stat} \pm 0.21_{syst}$ ns = $50.56 \pm 0.25$ 
ns \cite{bib:tauL_prelim} compatible with previous KLOE measurements.
The statistical error can be improved by decreasing the lower limit of the fit region, properly
accounting for the $K_L$ beam losses on the regenerating surfaces; 
the statistical error on the $K_L$ lifetime is expected to decrease to $\sim$0.1 ns. The
systematic error arising from the tagging efficiency, due only to detector acceptance, is expected to
decrease also.

\subsection{$K_S$ lifetime}
KLOE measures the $K_S$ lifetime ($\tau_S$) with a pure $K_S$ beam and event-by-event knowledge of the $K_S$
momentum. 
$\tau_S$ can be measured as a function of sidereal time which is interesting for tests
of quantum mechanics, CPT and Lorentz invariance~\cite{sidereal}. 
The lifetime is obtained by fitting the proper time, t$^\star$, distribution of $K_S\to\pi^+\pi^-$ decays. The 
resolution after event reconstruction 
is not sufficient for obtaining a lifetime accuracy of 0.1\%. 
The t$^\star$ resolution improves 
by reconstructing the IP event-by-event using a
geometrical fit, selecting events with pions decaying at large angle with respect to the $K_S$ path, and
rejecting poorly measured tracks by a cut on the the track fit $\chi^2$ value. 
The efficiency of this selection is $\sim$13\%. 
Since the resolution depends on the $K_S$ direction, we fit to the proper time distribution from -2 to 7 $\tau_S$
for each of 270 bins in $cos(\theta_K)$ and $\phi_K$.
The statistical error on $\tau_S$ is less than 0.1\%. 
With the full KLOE statistics the preliminary result $\tau_S
=  (89.56\pm 0.03\pm 0.07)\;\mathrm{ps}$~\cite{bib:tauS_prelim}
has been obtained, with the aim of reaching $\sim$0.03 ps final systematic uncertainty. 
A relative error of 0.03\% on $\tau_S$ is expected scaling this result to the 
KLOE-2 data sample.

\section{Measurement of $R_K=\Gamma(Ke2)/\Gamma(K\mu2)$}
\label{Kedue}
The SM prediction of $R_K$ benefits from cancellation of hadronic uncertainties 
to a large extent and therefore can be calculated with high precision. Including radiative
corrections, the total uncertainty is less than 0.5 per mil \cite{Marciano:1987ja,Cirigliano:2007xi}:
\begin{equation}
\label{eq:rksm}
 R_K = (2.477\pm0.001)\times 10^{-5}{\mbox .}
\end{equation}
Since the electronic channel is helicity-suppressed by the V$-$A structure of the charged weak
current, $R_K$ can receive contributions from physics beyond the SM, for example from multi-Higgs
effects inducing an effective pseudoscalar interaction. It has been shown in 
Ref. \cite{Masiero:2005wr} that deviations
from the SM of up to few percent on $R_K$ are quite possible in minimal supersymmetric extensions
of the SM and in particular should be dominated by lepton-flavor violating contributions with
tauonic neutrinos emitted in the electron channel:
\begin{equation}
\label{eq:rkbsm}
 R_K = R_K^{SM}\times \left[1+\frac{m_K^4}{m_H^4}\frac{m_\tau^2}{m_e^2}|\Delta_R^{31}|^2tan^6\beta \right]{\mbox ,}
\end{equation}
where $M_H$ is the charged-Higgs mass, $\Delta_R^{31}$ is the effective $e-\tau$ 
coupling constant depending on MSSM parameters, and $tan\beta$ is the ratio of the two vacuum expectation values.
Note that the pseudoscalar constant $f_K$ cancels in $R_K^{SM}$.
In order to compare with the SM prediction at this level of accuracy, one has to treat carefully
the effect of radiative corrections, which contribute to nearly half the $K_{e2\gamma}$ width. In particular,
the SM prediction of Eq. \ref{eq:rkbsm} is made considering all photons emitted by the process of internal
bremsstrahlung (IB) while ignoring any contribution from structure-dependent direct emission
(DE). Of course both processes contribute, so in the analysis DE is considered as a background
which can be distinguished from the IB width by means of a different photon energy spectrum.

Using the present KLOE dataset collected at the $\phi$ peak, and corresponding to $\sim$3.6 billion
$K^+K^-$ pairs, a measurement of $R_K$ with an accuracy of about 1 \% has been performed.
$\phi$ mesons are produced, essentially at rest and decay into $K^+K^-$
pairs with a BR of $\sim$49\%.
Kaons get a momentum of $\sim$100 \MeV\ which translates into a low speed, $\beta_K\sim0.2$. 
$K^+$ and $K^-$ decay with a mean length of $\lambda_\pm\sim$90 cm and can be distinguished from their decays
in flight to one of the two-body final states $\mu\nu$ or $\pi\pi^0$.
The kaon pairs from $\phi$ decay are produced in a pure J$^{PC}$ = 1$^{--}$ quantum state, so that
observation of a $K^+$ in an event signals, or tags, the presence of a $K^-$ and vice versa; highly
pure and nearly monochromatic $K^\pm$ beams can thus be obtained and exploited to achieve high
precision in the measurement of absolute BRs.
KLOE DC constitutes a fiducial volume for $K^\pm$ decays extending for $\sim1\lambda_\pm$. 
The momentum resolution for tracks at large polar angle is $\sigma_p/p\le0.4$\%. 
The c.m. momenta reconstructed from
identification of 1-prong $K^\pm\to\mu\nu,\pi\pi^0$ decay vertices in the DC peak around the expected
values with a resolution of 1–1.5 \MeV, thus allowing clean and efficient $K^\pm$ tagging.
Given the $K^\pm$ decay length, the selection of one-prong $K^\pm$ decays in the DC
required to tag $K^\mp$ has an efficiency smaller than 50\%. In order to keep the statistical uncertainty
on the number of $K\to e\nu$ counts below 1\%, a ``direct search'' for $K\to e\nu$ and $K\to\mu\nu$ decays
is performed, without tagging. Since the wanted observable is a ratio of BRs for two channels
with similar topology and kinematics, one expects to benefit from some cancellation of the
uncertainties on tracking, vertexing, and kinematic identification efficiencies. Small deviations
in the efficiency due to the different masses of $e$'s and $\mu$'s will be evaluated using MC.
 Selection starts by requiring a kaon track decaying in a DC fiducial volume (FV) with laboratory 
momentum between 70 and 130 \MeV, and a secondary track of relatively high momentum
(above 180 \MeV). The FV is defined as a cylinder parallel to the beam axis with length of 80 cm,
and inner and outer radii of 40 and 150 cm, respectively. Quality cuts are applied to ensure
good track fits.
    A powerful kinematic variable used to distinguish $K\to e\nu$ and $K\to\mu\nu$ decays from the
background is calculated from the track momenta of the kaon and the secondary particle: 
assuming $m_\nu=0$, the squared mass of the secondary particle ($m_\ell^2$) is evaluated. 
The selection applied is enough for clean identification of a $K\to\mu\nu$ sample, 
while further rejection is needed in order to identify $K\to e\nu$ events: 
the background, which is dominated by badly reconstructed $K\to\mu\nu$ events, 
is $\sim$10 times more frequent than the signal in the region around $m_e^2$.

\begin{figure}[ht]\centering
  \includegraphics[width=0.35\linewidth]{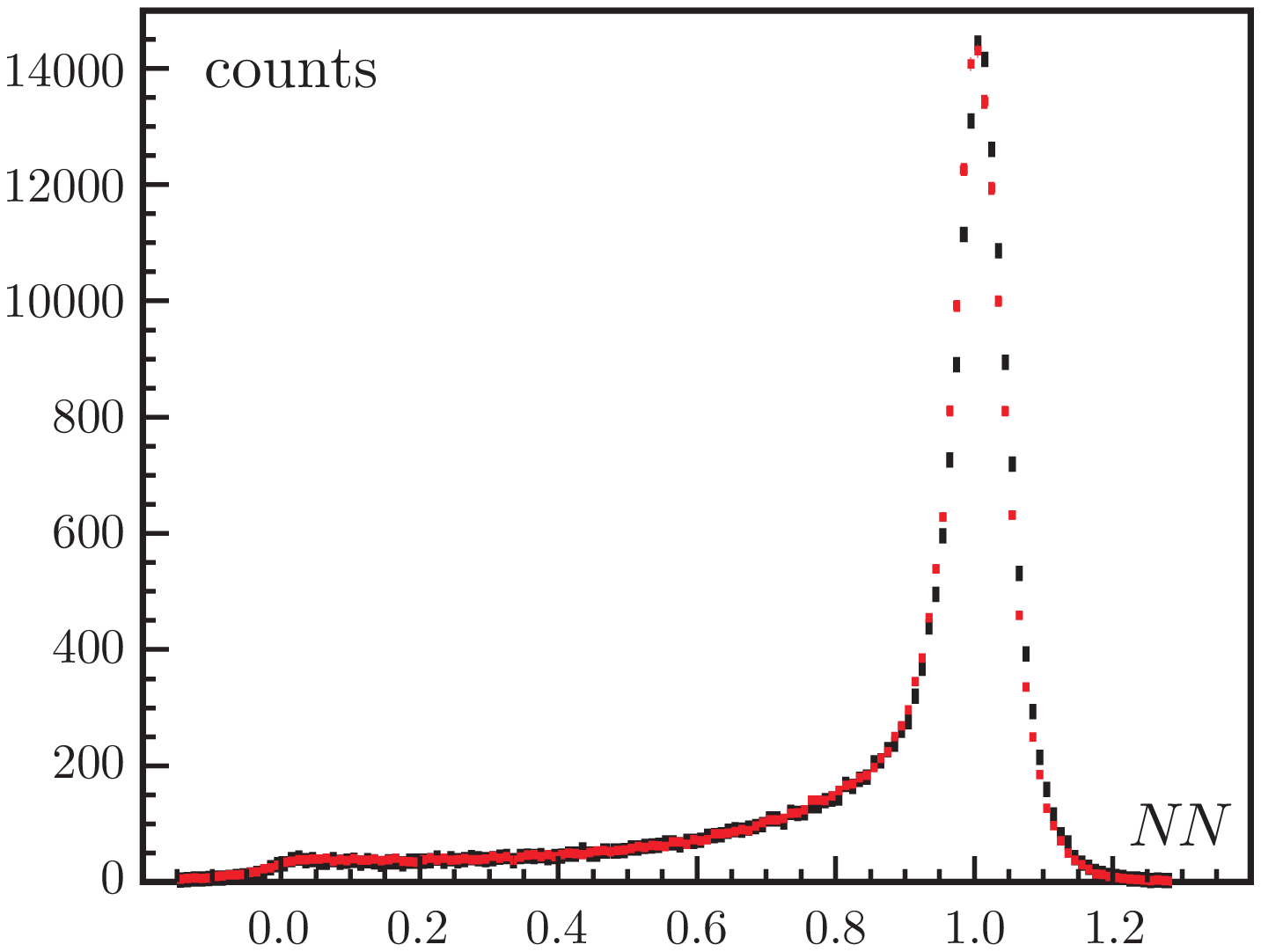}
  \includegraphics[width=0.3\linewidth]{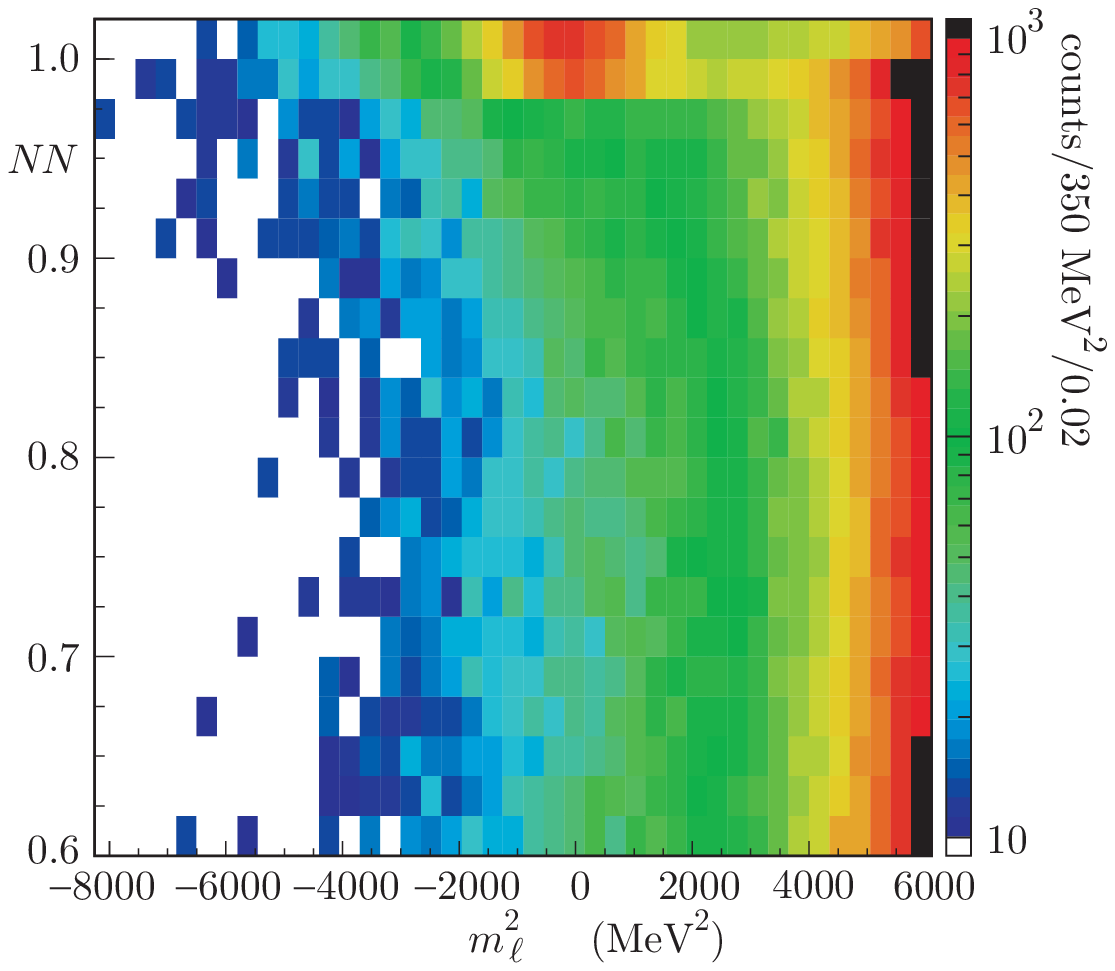}
  \caption{Left: Neural-network output,\nn, for electrons of a $K_L\to \pi e \nu$
   sample from data (black) and MC (red).
   Right: Data density in the \nn, $m_\ell^2$ plane: the signal is clearly visible
   at $m_\ell^2\sim0$ and $\nn\sim1$.}
  \label{ke2:pidNN}
\end{figure}

Information from the EMC is used to improve background rejection. To this purpose, we
extrapolate the secondary track to the EMC surface and associate it to a nearby EMC cluster.
For electrons, the associated cluster is close to the EMC surface and the cluster energy $E_{cl}$ is
a measurement of the particle momentum $p_{ext}$ , so that $E_{cl}/p_{ext}$ peaks around 1. For muons,
clusters tend to be more in depth in the EMC and $E_{cl}/p_{ext}$ tends to be smaller than 1, since
only the kinetic energy is visible in the EMC. Electron clusters can also be distinguished from
$\mu$ (or $\pi$) clusters, since electrons shower and deposit their energy mainly in the first plane of EMC,
while muons behave like minimum ionizing particles in the first plane and deposit a sizable
fraction of their kinetic energy from the third plane onward, when they are slowed down to rest
(Bragg’s peak). 
Particle identification has been therefore based on the
asymmetry of energy deposits between the first and the next-to-first planes, on the spread of
energy deposits on each plane, on the position of the plane with the maximum energy, and on the
asymmetry of energy deposits between the last and the next-to-last planes. All information are
combined with neural network (\nn) trained on $K_L\to\pi l\nu$ data, taking into account variations
of the EMC response with momentum and impact angle on the calorimeter. The distribution of
the \nn\ output for an independent $K_L\to\pi e\nu$ sample is shown in the left panel of 
Fig. \ref{ke2:pidNN} for data and Monte Carlo (MC). 
Additional separation has been obtained using time of flight information.
The number of $K\to e\nu(\gamma)$ is determined with a binned likelihood fit to the two-dimensional
\nn\ vs $m_\ell^2$ distribution. 
The data distribution of \nn\ as function of $m_\ell^2$ is
shown in Fig. \ref{ke2:pidNN} right. A clear \ke\ signal can be seen at $m_\ell^2\sim0$ and $\nn\sim1$.
Distribution shapes for signal and $K\mu2$ background, other sources being
negligible, are taken from MC; the normalization factors for the two components are the only fit
parameters. In the fit region, a small fraction of $K\to e\nu(\gamma)$ events is due to the direct-emission
structure-dependent component (DE): the value of this contamination, $f_{DE}$, is fixed in the fit to
the expectation from simulation. This assumption has been evaluated by performing a dedicated
measurement of DE, which yielded as a by-product a determination of $f_{DE}$ with a 4\% accuracy \cite{Ambrosino:2009rv}.
This implies a systematic error on Ke2 counts of 0.2\%, as obtained by repeating the fit with
values of $f_{DE}$ varied within its uncertainty.
In the fit region, we count 7064$\pm$102 $K^+\to e^+\nu(\gamma)$ and 
6750$\pm$101 $K^-\to e^-\nu(\gamma)$ events.

\begin{figure}[ht]\centering
    \includegraphics[width=0.3\textwidth]{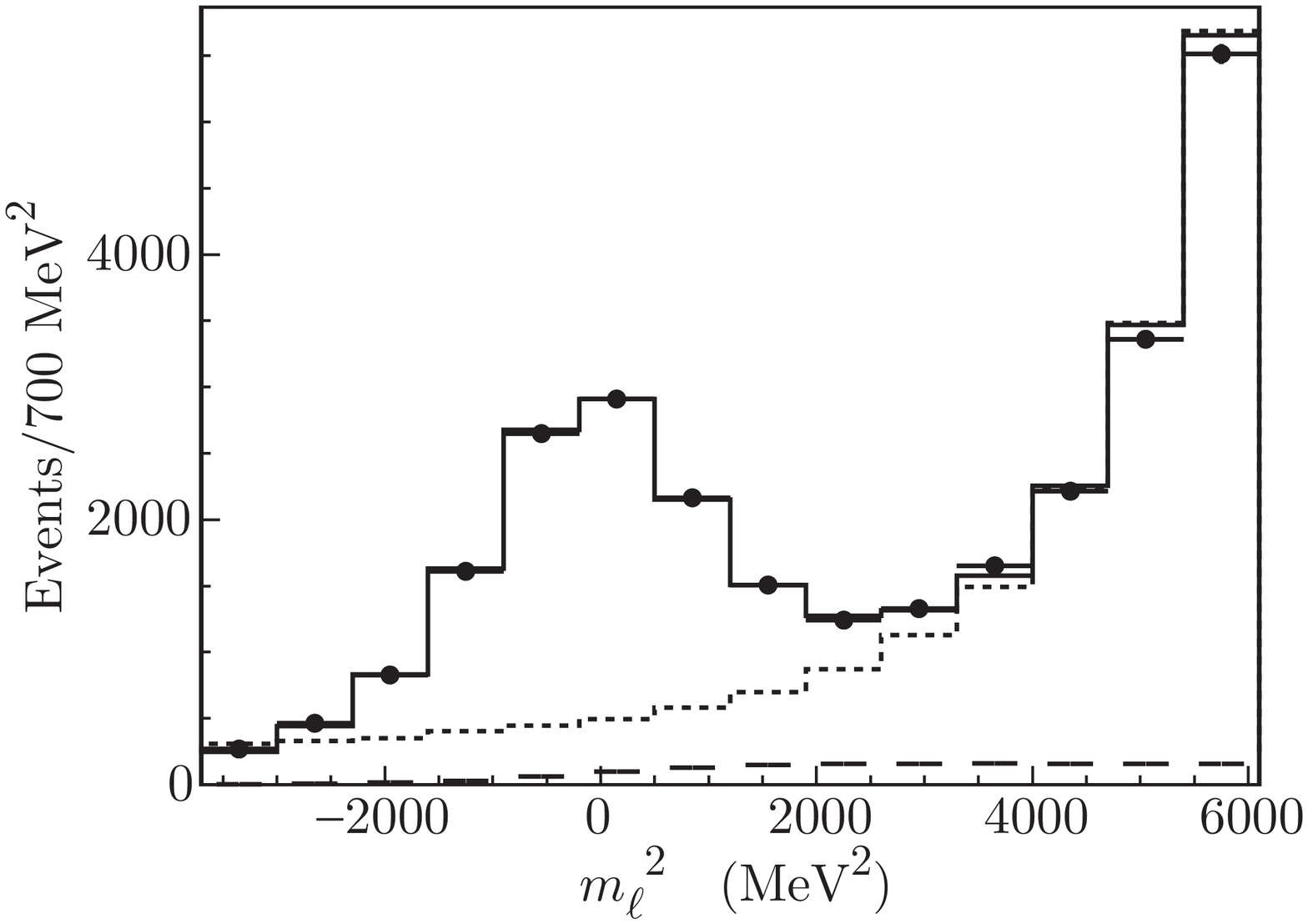}\kern1cm
    \includegraphics[width=0.3\textwidth]{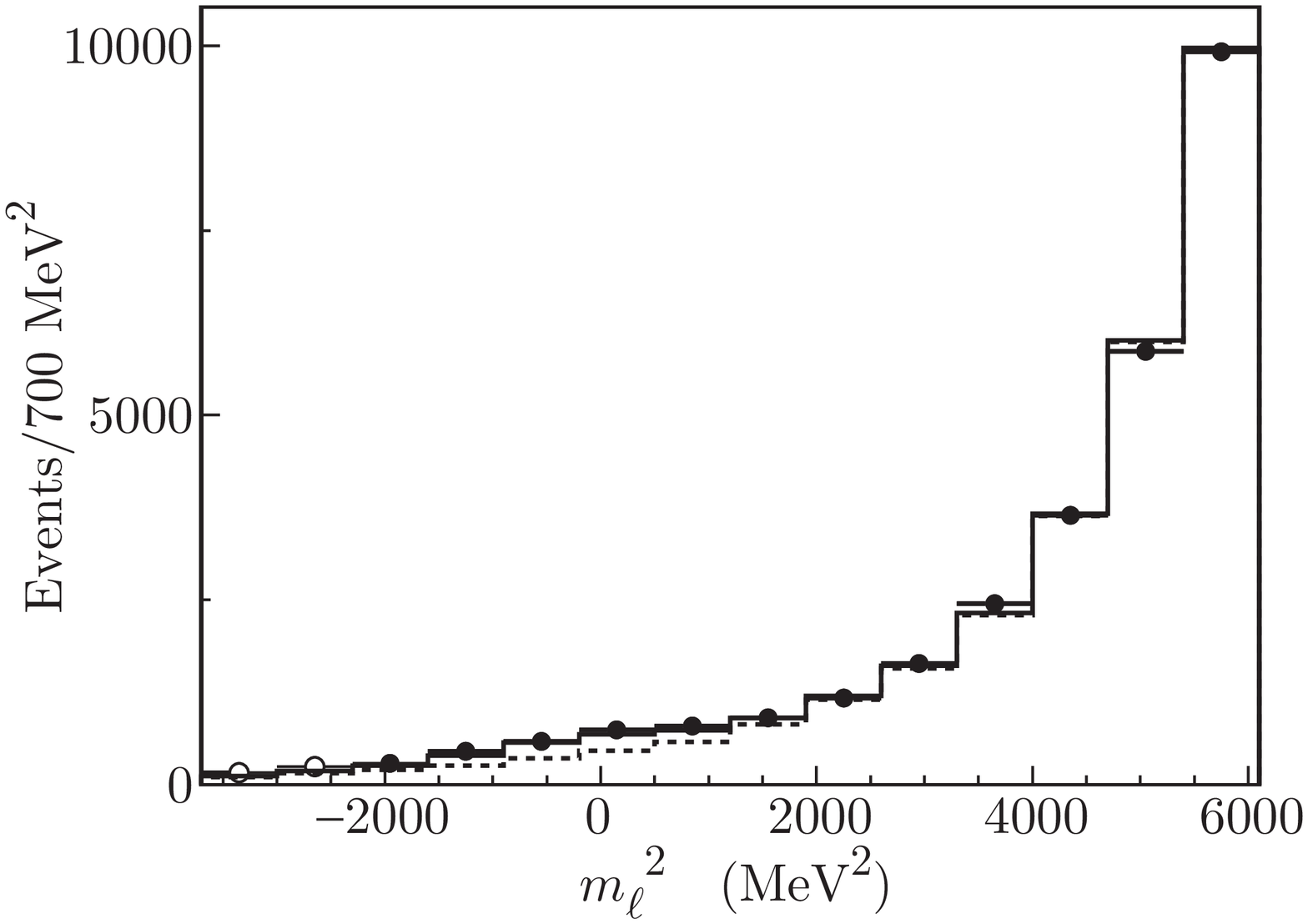}\kern1cm
   \includegraphics[width=0.25\linewidth]{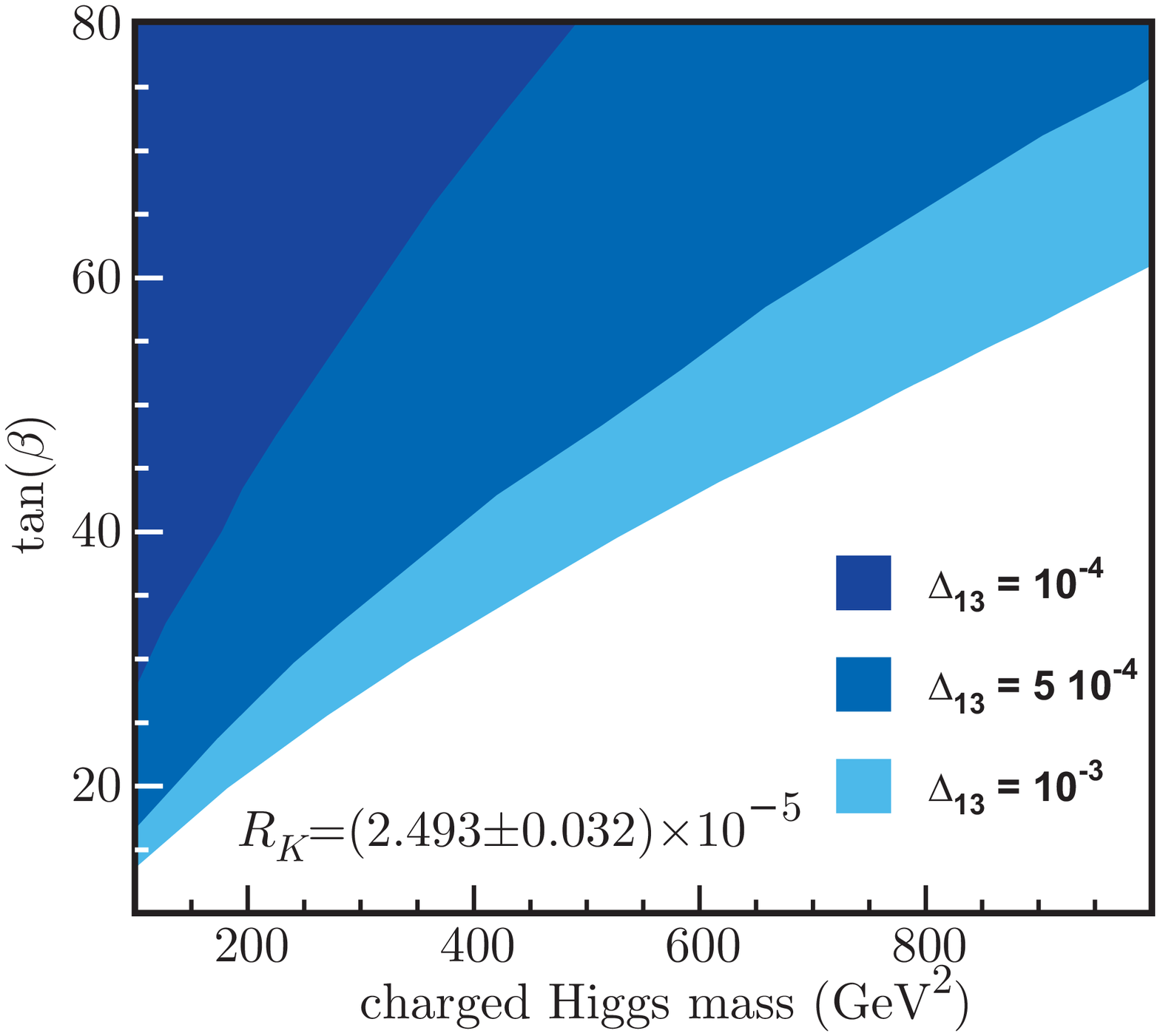}
  \caption{Fit projections onto the $m_\ell^2$ axis for $\nn>0.98$ (left) and $\nn<0.98$ (center), for data (black dots), MC fit (solid line), and  \kmd\ background (dotted line). The contribution from \ked\ events
with $E_\gamma>10$ MeV is visible in the left panel (dashed line). Right: Excluded regions at 95\% C.L. in the plane
 $M_H$--$\tan \beta$ for $\Delta_R^{31}=10^{-4}, 5\times 10^{-3},10^{-3}$.}
  \label{fig:fitke2}
\end{figure}

Fig. \ref{fig:fitke2} shows the sum of fit results for $K^+$ and $K^-$ projected onto the 
$m_\ell^2$ axis in a signal (\nn$>$0.98) and a background (\nn$<$0.98) enhanced region.
To assess the uncertainty on the $R_K$ measurement arising from limited knowledge of the
momentum resolution we have examined the agreement between the $m_\ell^2$ distributions for data
and MC in the $K\mu2$ region. For the \nn\ distribution, the EMC response at the cell level has
been tuned by comparing data and MC samples. In order to evaluate the systematic error
associated with these procedures, we studied the result variation with different fit range values,
corresponding to a change for the overall Ke2 purity from $\sim$75\% to $\sim$10\%. 
The results are stable within statistical fluctuations. A systematic uncertainty of 0.3\% for $R_K$ 
is derived ``\`a la PDG'' \cite{Amsler:2008zzb} by scaling the uncorrelated errors so that the 
reduced $\chi^2$ value of results is 1.

The number of $K\mu2$ events in the same data set is extracted from a fit to the $m_\ell^2$ 
distribution. The fraction of background events under the muon peak is estimated from MC to be $<$0.1\%.
We count $2.878\times10^8 (2.742\times10^8) K^+_\mu2 (K^-_\mu2)$ events. 
Difference in $K^+$ and $K^-$ counting is
ascribed to $K^-$ nuclear interactions in the material traversed.

The ratio of Ke2 to $K\mu2$ efficiency is evaluated with MC and corrected for data-to-MC ratios
using control samples. To check the corrections applied we also measured $R_3$ = BR(Ke3)/BR(K$\mu$3),
in the same data sample and by using the same methods for the evaluation of the efficiency as
for the $R_K$ analysis. We found $R_3$ = 1.507(5) and $R_3$ = 1.510(6), for $K^+$ and $K^-$ respectively.
These are in agreement within a remarkable accuracy with the expectation from 
world-average~\cite{Antonelli:2008jg} form-factor slope measurements, $R_3$ = 1.506(3).

The final result is $R_K = (2.493\pm0.025\pm0.019)\times10^{-5}$  \cite{Ambrosino:2009rv}. 
The 1.1\% fractional statistical error
has contributions from signal count fluctuation (0.85\%) and background subtraction. The 0.8\%
systematic error has a relevant contribution (0.6\%) from the statistics of the control samples
used to evaluate corrections to the MC. The result does not depend on K charge: quoting only
the uncorrelated errors, $R_K$($K^+$) = 2.496(37)10$^{-5}$ and $R_K$($K^-$) = 2.490(38)10$^{-5}$.
The result in agreement with SM prediction of Eq. \ref{eq:rksm}. Including the new KLOE result, the
world average reaches an accuracy at the \% level: $R_K$ = 2.468(25)$\times10^{-5}$. In the framework of
MSSM with LFV couplings, the $R_K$ value can be used to set constraints in the space of relevant
parameters (see Eq. \ref{eq:rkbsm}). The regions excluded at 95\% C.L. in the plane tan$\beta$–charged Higgs mass
are shown in the right panel of Fig. \ref{fig:fitke2} 
for different values of the effective LFV coupling $\Delta_R^{31}$

\section{Conclusions}
The experimental precision in leptonic and semileptonic kaon decays is nicely
matched below the percent level by theoretical precision, allowing to perform very precise 
measurements of SM parameters and to set  stringent bounds on physics beyond the SM.
KLOE contributed with the most comprehensive set of results from a single experiment, 
giving a fundamental contribution to the 0.2\% world accuracy on the determination of \Vusf.
KLOE result on $R_K$ improves the accuracy with which it is known by a factor of 5 with respect to the
present world average and allows severe constraints to be set
on new physics contributions in the MSSM with lepton flavor violating couplings.

\acknowledgments

\end{document}
\endinput